\documentclass[12pt]{article}
\newcommand{\be}{\begin{equation}}
\newcommand{\ee}{\end{equation}}

\newcommand{\bi}{\begin{itemize}}
\newcommand{\ei}{\end{itemize}}
\newcommand{\bea}{\begin{eqnarray}}
\newcommand{\eea}{\end{eqnarray}}
\newcommand{\ba}{\begin{array}}
\newcommand{\ea}{\end{array}}

\usepackage[left=2.50cm, right=2.50cm, top=2.50cm, bottom=2.50cm]{geometry}
\usepackage[utf8]{inputenc}
\usepackage{cancel}
\usepackage[english]{babel}
\usepackage{physics}
\usepackage{hyperref}
\usepackage{amsthm}
\usepackage{mathtools}
\usepackage{graphicx}
\usepackage{changepage}
\usepackage{amssymb,amsmath}
\usepackage{verbatim}
\usepackage{empheq}
\usepackage{xcolor}
\usepackage{tikz}
\usepackage{float}
\usepackage{multirow}
\usepackage{multicol}
\usepackage{amsmath}
\usepackage{centernot}
\usepackage[hang,flushmargin]{footmisc} 
\usepackage[normalem]{ulem}
\usetikzlibrary{tikzmark}
\numberwithin{equation}{section}
\hypersetup{hidelinks}

\newlength{\bibitemsep}
\newlength{\bibparskip}
\let\oldthebibliography\thebibliography
\renewcommand\thebibliography[1]{%
  \oldthebibliography{#1}%
  \setlength{\parskip}{\bibitemsep}%
  \setlength{\itemsep}{\bibparskip}%
}
\begin{document}
\par
\bigskip
\Large
\noindent
{\bf Maxwell theory of fractons \\
\par
\rm
\normalsize

\hrule

\vspace{1cm}

\large
\noindent
{\bf Erica Bertolini$^{1,2,a}$}, 
{\bf Nicola Maggiore$^{1,2,b}$}\\

\par
\small
\noindent$^1$ Dipartimento di Fisica, Universit\`a di Genova, Italy.
\smallskip

\noindent$^2$ Istituto Nazionale di Fisica Nucleare - Sezione di Genova, Italy.

\smallskip

\vspace{1cm}

\noindent
{\tt Abstract}

We show that the main properties of the fracton quasiparticles can be derived from a generalized covariant Maxwell-like action. Starting from a rank-2 symmetric tensor field $A_{\mu\nu}(x)$, we build a partially symmetric rank-3 tensor field strength $F_{\mu\nu\rho}(x)$ which obeys a kind of Bianchi identity. The most general action invariant under the covariant ``fracton'' transformation $\delta_{fract}A_{\mu\nu}(x)=\partial_\mu\partial_\nu\Lambda(x)$ consists of two independent terms: one describing Linearized Gravity (LG) and the other referable to fractons. The whole action can be written in terms of $F_{\mu\nu\rho}(x)$, and the fracton part of the invariant Lagrangian writes as $F^2(x)$, in analogy with Maxwell theory. The canonical momentum derived from the fracton Lagrangian coincides with the tensor electric field appearing in the fracton Literature, and
the field equations of motion, which have the same form as the covariant Maxwell equations ($\partial^\mu F_{\alpha\beta\mu}(x)=0$), can be written in terms of the generalized electric and magnetic fields and yield two of the four Maxwell equations (generalized electric Gauss and Amp\`ere laws), while the other two (generalized magnetic Gauss and Faraday laws) are consequences of the  ``Bianchi identity'' for the tensor $F_{\mu\nu\rho}(x)$, as in Maxwell theory. In the covariant generalization of the fracton theory, the equations describing the fracton limited mobility, $i.e.$ the charge and dipole conservation, are not external constraints, but rather
consequences of the field equations of motion, hence of the invariant action and, ultimately, of the fracton covariant symmetry. Finally, we increase the known analogies between LG and fracton theory by noting that both satisfy the generalized Gauss constraint which underlies the limited mobility property, which one would  not expect in LG.


\vspace{\fill}

\noindent{\tt Keywords:} \\
Quantum field theory, fractons, Maxwell gauge field theory, symmetric tensor gauge field theory, linearized gravity.

\vspace{1cm}

\hrule
\noindent{\tt E-mail:
$^a$erica.bertolini@ge.infn.it,
$^b$nicola.maggiore@ge.infn.it.
}
\newpage

\section{Introduction, summary of results and discussion}

Fractons are quasiparticles with the defining property of having restricted mobility \cite{Nandkishore:2018sel,Pretko:2020cko,Chamon:2004lew,Haah:2011drr,Vijay:2016phm,Pretko:2016lgv,Pretko:2016kxt,Pretko:2017xar,rasmussen,Pretko:2018jbi,Seiberg:2020wsg}. In particular ``true'' fractons cannot move at all, while other quasiparticles which can be traced back to fractons can move only in a subdimensional space, like ``lineons'', which move on a line  and ``planons'', which move on a plane \cite{Pretko:2020cko,Seiberg:2020wsg}. 
The first observations of a fracton-like behaviour appeared in lattice spin models \cite{Chamon:2004lew}, and since then many developments followed \cite{Haah:2011drr,Vijay:2016phm,Yoshida:2013sqa,Vijay:2015mka,Ma:2017aog,Shirley:2017suz}. Lattice models describing fractons fall into two classes, depending on their particle content:
``type I'', the most representative of which is the X-cube model \cite{Vijay:2016phm}, has both fractons and 1,2-dimensional particles, while
``type II'', of which the Haah's code \cite{Haah:2011drr} is the prototypical example, describes fractons only.
Fractons, like Linearized Gravity (LG), can be described by a gauge theory of a symmetric tensor field, which generalizes the ordinary Maxwell electromagnetism for a vector field \cite{Pretko:2016lgv}. This
class of fracton theories, which can be related to the previous lattice models via a Higgs-like mechanism \cite{Ma:2018nhd,Bulmash:2018lid}, were first introduced to describe gravity-related phenomena \cite{Gu:2006vw,Xu:2006, Gu:2009jh,Xu:2010eg}, and later were developed 
into the actual fracton theory \cite{Pretko:2016lgv,Pretko:2016kxt,Pretko:2017xar,rasmussen}.
Written in terms of a rank-2 symmetric tensor field $A_{ij}(x)$ ($i,j...$ spatial indices), the typical starting point is a Maxwell-like Hamiltonian ($E^2(x)+B^2(x)$), where the ``electric'' field $E^{ij}(x)$ is the conjugate momentum of $A_{ij}(x)$ as in standard electromagnetism, and the ``magnetic'' field $B_{ij}(x)$ is defined as the gauge invariant object depending on the lowest possible number of derivatives of $A_{ij}(x)$ \cite{Pretko:2017xar}. From these definitions, generalized Maxwell equations follow  
\cite{Pretko:2016lgv}: Faraday's equation is a relation between $E_{ij}(x)$ as conjugate momentum and the time derivative of $B_{ij}(x)$, while Amp\`ere law is a Hamilton's equation for $E_{ij}(x)$. Finally, the usual Gauss theorem in this picture is not really a theorem, but, rather, is imposed as a constraint. 
The gauge transformation of $A_{ij}(x)$    is crucial since, besides determining $B_{ij}(x)$, it is strictly related to the Gauss-like constraint, which has a key role in implementing the restricted mobility of fractons \cite{Nandkishore:2018sel,Pretko:2016lgv,Pretko:2016kxt,Pretko:2017xar,rasmussen,Pretko:2018jbi}. There are two possibilities \cite{Nandkishore:2018sel,Pretko:2020cko,Pretko:2016lgv,Pretko:2016kxt,Pretko:2017xar,Pretko:2018jbi,Blasi:2022mbl}:
\begin{itemize}

\item {\bf scalar charge theory :} the Gauss constraint and the gauge transformation of $A_{ij}(x)$ are
\bea
\partial_i\partial_jE^{ij} &=&0 \label{dedeE=0}\\
\delta A_{ij} &=&\partial_i\partial_j\Lambda \label{deA=dedelambda}\ .
\eea
In the presence of fractonic matter one can define a charge density operator $\rho(x)$ \cite{Pretko:2018jbi} and the Gauss constraint generalizes to \cite{Pretko:2016kxt}
\be
\partial_i\partial_jE^{ij}=\rho\ .
\label{dedeE=rho}\ee
The restricted mobility becomes evident, since \eqref{dedeE=rho} implies charge and dipole \mbox{($p^i(x)=x^i\rho(x)$)} neutrality when integrated on an infinite volume $V$ (or up to boundary terms) \cite{Nandkishore:2018sel,Pretko:2020cko,Pretko:2016lgv,Pretko:2016kxt,Pretko:2017fbf}
\bea
\int dV\partial_i\partial_jE^{ij}=\int dV\rho &=&0 \label{constr1}\\
-\int dV\partial_iE^{ik}=\int dVx^k\rho=\int dVp^k &=&0\ .\label{dipoleneutr}
\eea
Eq. \eqref{dipoleneutr} states that
single charges cannot move (fractons) due to dipole conservation, while dipoles do.

\item {\bf vector charge theory :} the Gauss constraint and the gauge transformation of $A_{ij}(x)$ are
\bea
\partial_iE^{ij} &=&0\label{}\\
\delta A_{ij} &=& \partial_i\Lambda_j+\partial_j\Lambda_i\ . \label{}
\eea
As in the scalar case, a vector charge density operator $\rho^j(x)$ can be defined \cite{Pretko:2018jbi} and the generalized Gauss constraint is \cite{Pretko:2016kxt}
\be
\partial_iE^{ij}=\rho^j\ ,
\label{}\ee
which immediately implies \eqref{dedeE=rho} together with the conservation laws \eqref{constr1} and \eqref{dipoleneutr}, but yields a further mobility constraint due to conservation of angular momentum           
			\begin{equation*}
			\int dV\epsilon_{0ijk}x^i\rho^j=-\int dV\epsilon_{0ijk}E^{ij}=0\ .
			\end{equation*}
Hence, vector charges can move only along the spatial direction related to the charged vector.
\end{itemize}
In Maxwell theory the $A_0(x)$ component of the gauge field $A_\mu(x)$ is a multiplier enforcing the standard Gauss constraint $\vec\nabla\cdot\vec E(x)=0$. Following this, in fracton models the Gauss constraint \eqref{dedeE=0} is implemented by introducing a Lagrange multiplier, as done for instance in \cite{Pretko:2017xar}, sometimes called $A_0(x)$ to enhance the Maxwell analogy. This multiplier seems to have no relation with the $A_{ij}(x)$ tensor field and in addition, due to this ``by hand'' implementation, the Lagrangian acquires an inhomogeneous number of (spatial) derivatives \cite{Pretko:2017xar}. For these reasons, despite all the similarities we mentioned, while Maxwell theory has a natural covariant formulation, the construction of fracton models appear to be intrinsically non-covariant.\\

In this paper we show that the main results concerning fractons, in particular the existence of tensorial electric and magnetic fields, the Gauss constraints, the Maxwell-like Hamiltonian and the dipole response to ``electromagnetic'' fields through a ``Lorentz force'', to cite a few, are indeed  consequences of a $covariant$ action for a symmetric rank-2 tensor field $A_{\mu\nu}(x)$, invariant under the covariant extension of the fracton transformation \eqref{deA=dedelambda}
\be
\delta_{fract}A_{\mu\nu}=\partial_\mu\partial_\nu\Lambda\ ,
\label{fractonsymintro}\ee
which therefore plays, as usual in quantum field theory, a central role. We shall show also that from the gauge tensor field $A_{\mu\nu}(x)$ it is possible to construct a rank-3 tensor $F_{\mu\nu\rho}(x)$ which we may call the fracton field strength, invariant under \eqref{fractonsymintro} and satisfying a kind of geometrical Bianchi identity. Quite surprisingly, the fracton action can be written in terms of the fracton field strength as $\int F^2$, as the ordinary Maxwell theory, and all the above mentioned equations characterizing fractons are nothing else than the ``Maxwell'' equations, without need of introducing any external constraint or particular request, and therefore are just consequences of the covariant symmetry \eqref{fractonsymintro}. Moreover, electric and magnetic tensor fields emerge naturally, and
in terms of these the action and the energy density read, respectively, $\int (E^2 - B^2)$ and $(E^2 + B^2)$. Finally, the Lorentz force for fracton dipoles derived ``by intuition'' ($sic$) in \cite{Pretko:2016lgv} is here recovered as part of the conservation law for the stress-energy tensor. As a matter of fact, the covariant generalization described in this paper makes apparent that the fracton theory is, indeed, a direct extension of the standard electromagnetic theory which can be formulated covariantly according to the typical field theory chain
$$\begin{array}{ccccc}
\mbox{symmetry} &\rightarrow& \mbox{action} &\rightarrow& \mbox{equations of motion} \\
\delta_{fract}A_{\mu\nu}=\partial_\mu\partial_\nu\Lambda(x)  
&\rightarrow&
-\frac{1}{6}\int d^4x\;F^{\mu\nu\rho}F_{\mu\nu\rho}
&\rightarrow&
\partial_\mu F^{\alpha\beta\mu}=0\ ,
\end{array}
$$
which really appears as a higher rank extension of Maxwell theory.\\

The relation between fractons and gravitons has been already remarked \cite{Nandkishore:2018sel,Pretko:2020cko,Blasi:2022mbl,Pretko:2017fbf}. From the field theory point of view this is evident from the covariant extension \eqref{fractonsymintro} of the fracton symmetry, which is a particular case of the stronger infinitesimal diffeomorphism transformation \cite{Blasi:2022mbl}
\be
\delta_{diff}A_{\mu\nu}=\partial_\mu\Lambda_\nu+\partial_\nu\Lambda_\mu\ .
\label{diffintro}\ee
In practice, this means that, while the diff symmetry \eqref{diffintro} uniquely defines the LG action, the most general action invariant under \eqref{fractonsymintro} is formed by two separately invariant terms: the LG action and the fracton action $\int F^2$, which is quite peculiar since, to our knowledge, this is the only case of a Lorentz invariant action which, although free and quadratic, shows a dimensionless constant which cannot be eliminated by a field redefinition, and which cannot be identified as a physical mass, like in topologically massive 3D gauge theory \cite{Deser:1981wh}. Coherently with this covariant picture, both fractons and LG can be given an electromagnetic formulation (the first, as discussed, in terms of tensors, while the second, known as gravitoelectromagnetism \cite{Mashhoon:2003ax,Carroll:2004st,Chatzistavrakidis:2020wum}, involves vectors), but, as we shall see, they also share
 the ``Gauss'' constraint \eqref{dedeE=0} (which is not an external constraint in our formalism) which underlies the fracton limited mobility property.\\

The analogy with ordinary electromagnetism can be pushed further through the 
topological $\theta$-term that can be added to the Maxwell Lagrangian (see for instance \cite{tong} and references therein).
The role of an analogous boundary term has been studied in the case of fractons in \cite{Pretko:2017xar} and, as for dyons, the result is that  the ``electric'' charge gains an additional contribution related to a ``magnetic'' vector charge \cite{Nandkishore:2018sel,Pretko:2020cko,Pretko:2017xar}. As the standard Witten effect has consequences in condensed matter physics, this higher rank version of the fracton $\theta$-term might give interesting results in the context of higher order topological phases \cite{You:2019bvu}. The case of a local, instead of constant,
$\theta(x)$ is relevant in axion models \cite{Peccei:1977hh,Peccei:1977ur}, where Maxwell equations acquire additional contributions \cite{Sikivie:1983ip,Wilczek:1987mv}. 
In \cite{Chatzistavrakidis:2020wum} a local $\theta(x)$-term has been added to LG, with mainly two consequences : 
a correction to the Newtonian gravitational field and a Witten-like effect for gravitational dyons, in which ``gravitipoles'' \cite{Zee:1985xqg} acquire mass. The results of \cite{Chatzistavrakidis:2020wum} for LG suggest the possibility of generalizing what has been found in \cite{Pretko:2017xar} for fractons, since both LG and fractons are described by a rank-2 symmetric tensor field.\\

There are of course, and fortunately, a few open questions, which deserve further efforts. The first is that we were not able to find a symmetry which separates fractons from gravitons. In other words: while the diff symmetry \eqref{diffintro} uniquely defines the LG action, the fracton symmetry \eqref{fractonsymintro} gives two invariant functionals. One must necessarily buy gravitons, together with fractons. The way out in field theory is to recover the fracton action for vanishing LG constant, but, still, it would be more satisfying to pick up the fracton action by means of an additional symmetry. Moreover: the fracton symmetry \eqref{fractonsymintro} (but also the original \eqref{deA=dedelambda}) is dimensionally problematic. In 4D the rank-2 tensor $A_{\mu\nu}(x)$ has mass dimension one, both in fracton and LG theory. Hence, to be dimensionally homogeneous, the fracton gauge transformation would require a scalar gauge parameter $\Lambda(x)$ with negative dimension
	\begin{equation}
	d=4\quad\Rightarrow\quad[A]=1\quad;\quad\delta A_{\mu\nu}=\partial_\mu\partial_\nu\Lambda\quad\Rightarrow[\Lambda]=-1\ ,
	\end{equation}
but this would not be the case in 6D, which would be the most ``natural'' spacetime dimensions for fractons to live in
	\begin{equation}
	d=6\quad\Rightarrow\quad[A]=2\quad;\quad\delta A_{\mu\nu}=\partial_\mu\partial_\nu\Lambda\quad\Rightarrow[\Lambda]=0\ .
	\end{equation}
 As we shall see, this reflects also in the fact that, in 4D, the stress-energy tensor is not traceless, hence the theory is not scale invariant, differently from the classical Maxwell theory. Instead, tracelessness is recovered in 6D. \\
Finally, we remark a subtle point which concerns the stress-energy tensor of the fracton action. In Section 3 we compute the stress-energy tensor, defined as
\be
T_{\alpha\beta}=-\frac{2}{\sqrt{-g}}\frac{\delta S}{\delta g^{\alpha\beta}}\ ,
\label{Tmunuintro}\ee
and we show that it has the correct components, which are exactly the higher rank generalizations of the Maxwell energy density ($T_{00}(x)$), of the Poynting vector ($T_{0i}(x)$) and of the stress tensor ($T_{ij}(x)$). The time component is conserved on shell, $i.e.\ \partial_\mu T^{\mu0}(x)=0,$ and gives the continuity equation relating the energy density to the Poynting vector. So far so good. But the space component of the stress-energy tensor conservation law is not exactly conserved. We find a breaking term which might be interpreted as follows. The stress-energy tensor \eqref{Tmunuintro} is the conserved current associated to the infinitesimal diff invariance \eqref{diffintro} \cite{Carroll:2004st}, {\it which is not} a symmetry of the theory defined by \eqref{fractonsymintro}. Hence, the stress-energy tensor \eqref{Tmunuintro} should not be conserved. Nonetheless, the fracton transformation \eqref{fractonsymintro} is a particular case of the general diff transformation \eqref{diffintro}. Hence, it is not that unexpected that the stress-energy tensor is $almost$ conserved. A further confirmation that $T_{\alpha\beta}(x)$ \eqref{Tmunuintro} is the correct one is that, when matter is added, quite remarkably the conservation equation gives exactly the Lorentz force for fracton dipoles that has been conjectured in \cite{Pretko:2016lgv}.\\

The paper is organized as follows. In Section 2 we derive the theory defined by the fracton symmetry \eqref{fractonsymintro}, composed by two terms, fracton and LG. We then construct the rank-3 fracton field strength $F_{\mu\nu\rho}(x)$ and we show that both the fracton and the LG actions can be written in terms of this tensor, which satisfies an identity analogous to the Bianchi one. We then compute the canonical momentum $\Pi^{\alpha\beta}(x)$ associated to $A_{\alpha\beta}(x)$ and derive the field equations of motion. In Section 3 we consider the fracton theory only, obtained by putting the LG constant to zero. Without imposing any external constraint, we recover the main properties of the fractons simply from the equations of motion, which, written in terms of the electric and magnetic tensor fields,  impressively reminds the ordinary Maxwell equations. We derive the stress-energy tensor and physically identify its components, which are the higher rank extensions of their Maxwell counterparts. We then write, interpret and discuss the stress-energy tensor conservation laws. In Section 4 we add matter to the theory, represented by a symmetric rank-2 tensor coupled to $A_{\alpha\beta}(x)$, and extend the previously found results in presence of matter. The most important achievement of this Section is the expression of the Lorentz force which describes how dipole fractons respond to the electromagnetic tensor fields. This Lorentz force coincides with the one conjectured in \cite{Pretko:2016lgv}. Finally, in Section 5 we add to the fracton action the generalized $\theta$-term, which, again, can be written both in terms of the electromagnetic tensor fields and of the fracton field strength, in complete analogy with Maxwell theory. We  recover and generalize previous results obtained in the context of LG \cite{Chatzistavrakidis:2020wum} and of fractons \cite{Pretko:2017xar} giving to the $\theta$-parameter a local dependence. Some final remarks can be found in Section 6.

\section{Fractons and linearized gravity}

\subsection{The symmetry}

    We adopt the standard point of view of field theory, that is to consider the symmetry as the birth certificate of a theory. In our case, the symmetry,  hereinafter ``fracton'' symmetry,  is the covariant generalization of the extended electromagnetic transformation \eqref{deA=dedelambda} invoked in \cite{Pretko:2016lgv,Pretko:2016kxt,Pretko:2017xar,rasmussen} for fractons, $i.e.$ 
\begin{equation}\label{dA}
	\delta_{fract} A_{\mu\nu}=\partial_\mu\partial_\nu\Lambda\ ,
\end{equation}
where $A_{\mu\nu}(x)$ is a rank-2 symmetric tensor field and $\Lambda(x)$ a local scalar parameter. 
The fracton transformation \eqref{dA} is obtained from the more general infinitesimal diffeomorphism transformation
\be\label{diff}
\delta_{diff} A_{\mu\nu}=\partial_\mu\Lambda_\nu + \partial_\nu\Lambda_\mu  
\ee
for the particular choice of the gauge parameter 
$\Lambda_\mu(x)=\frac{1}{2}\partial_\mu\Lambda(x)$.
The most general 4D action invariant under the fracton symmetry \eqref{dA} is a linear combination of two invariant actions
\be
S_{inv}=g_1S_{fract}+g_2S_{LG}\ ,
\label{Sinvg1g2}\ee
where
\bea
S_{fract} &=&
\int d^4x 
\left(\partial_\rho A_{\mu\nu}\partial^\rho A^{\mu\nu}- 
\partial_\rho A_{\mu\nu} \partial^\mu A^{\nu\rho} \right)\label{Sfract}\\
S_{LG} &=& \int d^4x\left(\partial_\mu A\partial^\mu A-\partial_\rho A_{\mu\nu}\partial^\rho A^{\mu\nu}-2\partial_\mu A\partial_\nu A^{\mu\nu}+2\partial_\rho A_{\mu\nu} \partial^\mu A^{\nu\rho}\right)\ , \label{SLG}
\eea
$g_1,g_2$ are dimensionless constants, and $A\equiv\eta^{\mu\nu}A_{\mu\nu}$ is the trace of the tensor field. 
The action $S_{LG}$ is readily recognized to be the linearized action for gravity \cite{Hinterbichler:2011tt}, while $S_{fract}$ is our candidate to be the covariant action for fractons, as we shall motivate in this paper. 
Hence, the space of 4D local integrated functionals invariant under the fracton symmetry \eqref{dA} has dimension two, and one of the two constants $g_1$ and $g_2$ can be reabsorbed by a redefinition of $A_{\mu\nu}(x)$, so that we have the rather peculiar feature that the free quadratic theory defined by the action \eqref{Sinvg1g2}, hence by the fracton transformation \eqref{dA}, depends on one constant. To our knowledge, this is the only example of a free quadratic covariant theory depending on a constant which cannot be reabsorbed by a field redefinition, without being identified as a mass, like in 3D topologically massive gauge theories \cite{Deser:1981wh}.
In particular we have that $S_{fract}$ \eqref{Sfract} and $S_{LG}$ \eqref{SLG} are both invariant under the fracton transformation \eqref{dA}
\be
\delta_{fract}S_{LG} = \delta_{fract}S_{fract} = 0\ ,
\label{}\ee
but only the LG action \eqref{SLG} is invariant under the diff transformation \eqref{diff}
\be
\delta_{diff}S_{LG}= 0\ ,
\label{}\ee
while the fracton $S_{fract}$ \eqref{Sfract}, hence the whole action $S_{inv}$ \eqref{Sinvg1g2}, is not
\be
\delta_{diff}S_{fract} = g_1\int d^4x \left[
2\partial^\mu\partial^\nu A_{\mu\nu}\partial^\rho\Lambda_\rho
-\partial^2A_{\mu\nu}(\partial^\mu\Lambda^\nu+\partial^\nu\Lambda^\mu)
\right]\neq0\ .
\label{notdiff}\ee

\subsection{The fracton field strength}

The first step towards a Maxwell theory for fractons, which is the  main purpose of this article, is the construction of the ``building block'' of the theory, namely the extension of the electromagnetic field strength $F_{\mu\nu}(x)$
\be
\begin{split}
A_\mu &\rightarrow F_{\mu\nu}=\partial_\mu A_\nu - \partial_\nu A_\mu  \\
A_{\mu\nu} &\rightarrow F_{\mu\nu\rho} =\quad ?
\end{split}
\label{}\ee
To this aim, we look for a rank-3 tensor built from the first derivative of the rank-2 tensor field $A_{\mu\nu}(x)$
\begin{equation}
F_{\mu\nu\rho}\equiv 
a_1\partial_\mu A_{\nu\rho}+a_2\partial_\rho A_{\mu\nu}+a_3\partial_\nu A_{\mu\rho}\ ,
	\label{Fmunurhoin}\end{equation}
where $a_i$ are dimensionless constants. As the electromagnetic tensor $F_{\mu\nu}(x)$ is invariant under the ordinary gauge transformation $\delta_{gauge}A_{\mu}(x)=\partial_\mu\Lambda(x)$, in the same way we require that $F_{\mu\nu\rho}(x)$ is invariant under the fracton symmetry \eqref{dA}, which gives a constraint on the coefficients $a_i$
\be
\delta_{fract} F_{\mu\nu\rho}=0\ \Rightarrow\ a_3=-(a_1+a_2)\ ,
\label{deltaF=0}\ee
so that
\begin{equation}\label{F}
F_{\mu\nu\rho}=
a_1\partial_\mu A_{\nu\rho}+a_2\partial_\rho A_{\mu\nu}-(a_1+a_2)\partial_\nu A_{\mu\rho}\ .
\end{equation}
As a consequence of its definition, the invariant tensor \eqref{F} has the properties listed in Table \ref{table1}, compared to those of the Maxwell field strength $F_{\mu\nu}(x)$.

\begin{center}
\begin{table}[ht]
\centering
  \begin{tabular}{ | l | c | c| }
    \hline
     & fractons & Maxwell \\ \hline
      invariance & $\delta_{fract}F_{\mu\nu\rho}=0$& $\delta_{gauge}F_{\mu\nu}=0$ \\ \hline
    cyclicity & $F_{\mu\nu\rho}+F_{\nu\rho\mu}+F_{\rho\mu\nu}=0$ & $F_{\mu\nu}+F_{\nu\mu}=0$ \\ \hline
    Bianchi & $\epsilon_{\alpha\mu\nu\rho}\partial^{\mu}F^{\beta\nu\rho}=0$ & 
   $ \epsilon_{\mu\nu\rho\sigma}\partial^\nu F^{\rho\sigma}=0$ \\
    \hline
  \end{tabular}
  \caption{\label{table1} Properties of the fracton and Maxwell field strengths.}
\end{table}
\end{center}
We remark that the fracton invariance of $F_{\mu\nu\rho}(x)$ \eqref{deltaF=0} and the property which we called ``cyclicity'' in Table \ref{table1} are equivalent
\begin{equation}
	\delta_{fract} F_{\mu\nu\rho}=0\quad\Leftrightarrow\quad F_{\mu\nu\rho}+F_{\nu\rho\mu}+F_{\rho\mu\nu}=0\quad\Leftrightarrow\quad a_1+a_2+a_3=0\ .
	\end{equation}	
All the physically relevant quantities (like for instance the equations of motion and the conjugate momenta) are obtained by making functional derivatives with respect to $A_{\mu\nu}(x)$, which is a symmetric tensor field. With the aim of writing everything in terms of the tensor field strength $F_{\mu\nu\rho}(x)$, it is natural to ask that also this latter is symmetric by the change of two indices, for instance the first two
\be
F_{\mu\nu\rho}=F_{\nu\mu\rho}
\label{Fmunu=Fnumu}\ ,\ee
which implies $a_2=-2a_1$\footnote{We checked that this is indeed the case, $i.e.$ $F_{\mu\nu\rho}(x)-F_{\nu\mu\rho}(x)$ is always ruled out.}.
Therefore, after a rescaling of $A_{\mu\nu}(x)$, our fracton field strength is
	\begin{equation}
	F_{\mu\nu\rho}=F_{\nu\mu\rho}=\partial_\mu A_{\nu\rho}+\partial_\nu A_{\mu\rho}-2\partial_\rho A_{\mu\nu}\ .
	\label{Fmunurho}\end{equation}
Rather surprisingly, the same symmetric tensor \eqref{Fmunurho} appears as an unnumbered comment in the final part of a 1988 paper by Y.S. Wu and A. Zee \cite{Wu:1988py} as a consequence of the covariant symmetry \eqref{dA}, but in a completely different context, since fractons were not even conceived yet.\footnote{We thank Giandomenico Palumbo for this remark.}
    	
\subsection{The fracton and LG actions}

The actions \eqref{Sfract} and \eqref{SLG} can be written in terms of the fracton field strength $F_{\mu\nu\rho}(x)$ \eqref{Fmunurho} as  
\bea
S_{fract} &=&
\frac{1}{6}\;\int d^4x\;F^{\mu\nu\rho}F_{\mu\nu\rho}
\label{SfractF}\\
S_{LG} &=& 
\int d^4x\; \left(
\frac{1}{4}F^\mu_{\ \mu\nu} F_\rho^{\ \rho\nu}-\frac{1}{6}F^{\mu\nu\rho}F_{\mu\nu\rho}
\right)\ .
\label{SLGF}
\eea
Notice that also the LG action \eqref{SLG} can be written in terms of the newly introduced tensor $F_{\mu\nu\rho}(x)$ \eqref{Fmunurho}. 
The fact that the fractonic component of the total action $S_{inv}$ \eqref{Sinvg1g2} turns out to be of the form $\int F^2$ tells us that we are on the right way to build a Maxwell theory of fractons, but the analogies are even more surprising in what follows.
\subsection{The canonical momentum $\Pi^{\alpha\beta}(x)$}

In the theory of the fracton quasiparticles an important role is played by the momentum canonically conjugated to $A_{\mu\nu}(x)$ \cite{Nandkishore:2018sel,Pretko:2016lgv,Pretko:2016kxt,Pretko:2017xar,Du:2021pbc}. From \eqref{Sinvg1g2} we have
\be
\Pi^{\alpha\beta}(g_1,g_2)\equiv\frac{\partial\mathcal L_{inv}}{\partial(\partial_tA_{\alpha\beta})}
=
-g_1 F^{\alpha\beta0}
-g_2\left[
\eta^{\alpha\beta}F_{\lambda}^{\ \lambda0}-\frac{1}{2}\left(\eta^{0\alpha}F_{\lambda}^{\ \lambda\beta}+\eta^{0\beta}F_{\lambda}^{\ \lambda\alpha}\right) -F^{\alpha\beta0}\right]\ ,
\label{E^ij inv}\ee
whose components are
\bea
	\Pi^{00}&=&0 \label{Pi00=0}\\
	\Pi^{i0}&=&-g_1F^{i00}-\frac{1}{2}g_2 F_j^{\ ji}\label{Pi0i}\\
	\Pi^{ij}&=&-g_1F^{ij0} +g_2(F^{ij0}-\eta^{ij}F_k^{\ k0}) \label{Piij}\ .
\eea
From \eqref{Pi00=0} we see that $A_{00}(x)$ is not a dynamical field for the whole theory (both fractons and LG). For what concerns LG alone, it is known \cite{Pretko:2017fbf,Hinterbichler:2011tt} that the components with a time index, $A_{00}(x)$ and $A_{0i}(x)$, have non-dynamical equations of motion, acting as Lagrange multipliers to enforce gauge
constraints, in the same way as $A_0(x)$ acts as a Lagrange multiplier enforcing Gauss law in Maxwell theory. The physical degrees of freedom are contained in the spatial symmetric tensor $A_{ij}(x)$. We shall see that this property concerning LG holds for fracton theory too, which therefore remarkably shares close similarities with both LG and Maxwell theory. We finally notice that for a particular combination of $g_1$ and $g_2$ the trace of $\Pi^{\alpha\beta}$ vanishes
\be
\eta_{\alpha\beta}\Pi^{\alpha\beta}
=\Pi^\alpha_{\ \alpha}=\Pi^i_{\ i}=-(g_1+2g_2)F_{\lambda}^{\ \lambda0}=0\quad \mbox{if $g_1+2g_2=0$}\ .
\label{tracelessconstr}\ee
This corresponds to the fact that, as already remarked in \cite{Blasi:2022mbl}, the theory defined by $S_{inv}$ \eqref{Sinvg1g2} at \eqref{tracelessconstr} does not depend on the trace of the tensor field $A_{\mu\nu}(x)$, further lowering the  number of degrees of freedom.

\subsection{The field equations of motion}

As the fracton and LG actions \eqref{SfractF} and \eqref{SLGF}, the field Equations of Motion (EoM) can be written in terms of the fracton field strength $F_{\mu\nu\rho}(x)$ as well
\bea
\frac {\delta S_{inv}}{\delta A^{\alpha\beta}} &=&
2g_1\left[
\left(\partial^\mu\partial_\alpha A_{\mu\beta}+\partial^\mu\partial_\beta A_{\mu\alpha}\right)
-\partial^2 A_{\alpha\beta}
\right]\nonumber\\
&&+2g_2\left[
\eta_{\alpha\beta}\left(\partial_\mu\partial_\nu A^{\mu\nu}-\partial^2A\right)
+\partial_\alpha\partial_\beta A
+ \partial^2 A_{\alpha\beta}
-2 \left(\partial^\mu\partial_\alpha A_{\mu\beta}+\partial^\mu\partial_\beta A_{\mu\alpha}\right)\right]\nonumber \\
	&=&
g_1\partial^{\mu}F_{\alpha\beta\mu}
+g_2\left[\eta_{\alpha\beta}\partial_\mu F_\nu^{\ \nu\mu}-\frac{1}{2}\left(\partial_\alpha F^\mu_{\ \mu\beta}+\partial_\beta F^{\mu}_{\ \mu\alpha}\right)-\partial^{\mu}F_{\alpha\beta\mu}\right]=0\ ,
\label{eom inv}
\eea
whose components are
\begin{itemize}
\item $\alpha=\beta=0$
\be
g_1\partial_iF^{00i} - g_2 \partial_i (F_\lambda^{\ \lambda i}+F^{00i}) =
2\partial_i (-g_1F^{i00}-\frac{1}{2}g_2F_j^{\ ji}) =
2\partial_i\Pi^{i0} = 0\ ,
\label{eomtot00}\ee
where we used $F^{00i}=-2F^{i00}$ and the definition of the canonical momentum $\Pi^{i0}$ \eqref{Pi0i}. 
\item $\alpha=0$, $\beta=i$
\be
g_1\partial_\lambda F^{0i\lambda}
-\frac{1}{2}g_2 
(\partial^0F_\lambda^{\ \lambda i} + \partial^iF_\lambda^{\ \lambda 0} + 2 \partial_\lambda F^{0i\lambda}) =
-\partial_0\Pi^{i0} 
+ g_1 \partial_j F^{0ij} 
-\frac{1}{2}g_2 (\partial^i F_\lambda^{\ \lambda 0} + 2\partial_jF^{0ij}) 
=0\;;
\label{eomtot0i}\ee
\item $\alpha=i$, $\beta=j$
\be
g_1\partial^\mu F_{ij\mu} + g_2[
\eta_{ij}\partial_\mu F_\nu^{\ \nu\mu}-\frac{1}{2}(\partial_iF^\mu_{\ \mu j} +\partial_jF^\mu_{\ \mu i})-\partial^\mu F_{ij\mu}]=0\ .
\label{eomtotij}\ee
\end{itemize}

\section{Maxwell theory for fractons}

In this Section we treat the case $g_2=0$, and we shall recover the main features generally attributed to the fracton quasiparticles \cite{Nandkishore:2018sel,Pretko:2016lgv,Pretko:2016kxt,Pretko:2017xar}, thus allowing us to justify the identification of $S_{fract}$ \eqref{SfractF} as the action for fractons.

\subsection{Electric/magnetic tensor fields and ``Maxwell'' equations}

As far as only fractons are considered, in \cite{Nandkishore:2018sel,Pretko:2016lgv,Pretko:2016kxt,Pretko:2017xar,Du:2021pbc} an electric tensor field $E^{ij}(x)$ is defined as spatial ``canonical momentum'' as follows
\begin{equation}
		E_{ij}\propto -\partial_tA_{ij}+\partial_i\partial_j A_0\ .
\label{electricfielddefPretko}\end{equation}	
We would like to show here that $E_{ij}(x)$ \eqref{electricfielddefPretko} can indeed be derived from the action \eqref{SfractF} in a way which also clarifies which is the origin of the scalar field $A_0(x)$ appearing in \eqref{electricfielddefPretko}. In fact, $A_0(x)$ cannot be directly part of a canonical momentum unless in the Lagrangian weird terms with three derivatives are admitted \cite{Pretko:2017xar}. The covariant extension \eqref{dA} 
of the fracton symmetry has a central role in determining \eqref{electricfielddefPretko}, without the need of any {\it ad-hoc} introduction. In fact, starting from the fracton transformation \eqref{dA} one gets the action $S_{inv}$ \eqref{Sinvg1g2} from which the spatial canonical momentum \eqref{Piij} is derived. In the case where only fractons are present, namely $g_2=0$, the canonical momentum $\left.\Pi^{ij}(x)\right|_{g_2=0}$ reads:
\begin{equation}\label{Pi^ij}
\left.\Pi^{ij}\right|_{g_2=0}= -g_1 F^{ij0} = g_1\left(2\partial^0A^{ij}-\partial^jA^{0i}-\partial^iA^{0j}\right)\ ,
\ee
which differs from \eqref{electricfielddefPretko}. Nevertheless, the  electric tensor field \eqref{electricfielddefPretko} can be indeed obtained from the spatial canonical momentum  $\left.\Pi^{ij}(x)\right|_{g_2=0}$ using the EoM \eqref{eom inv} with $g_2=0$, $i.e.$ those derived from the fracton action \eqref{SfractF} alone, which closely remind the usual Maxwell equations
\be
\partial^\mu F_{\alpha\beta\mu}=0\ .
\label{eom2}\ee
In fact, taking \eqref{eom2} at $\alpha=\beta=0$ (or, equivalently, \eqref{eomtot00} at $g_2=0$) we have 
\be
\partial^iF_{00i}=
2\partial^i\left(\partial_0A_{0i}-\partial_iA_{00}\right)=0\ .
\label{eom00F}\ee
A particular solution is given by 
\begin{equation}\label{A0}
		A_{0\mu}=A_{\mu0}\equiv\partial_\mu A_0\ ,
\end{equation}
which introduces the missing scalar potential $A_0(x)$. 
What renders remarkable the solution \eqref{A0}, which is a direct consequence of our covariant approach, is that it leads to recover, up to a constant,  the electric tensor field \eqref{electricfielddefPretko}. In fact, using \eqref{A0} in \eqref{Pi^ij} we get
\begin{equation}
(\Pi^{ij}|_{g_2=0})|_\eqref{A0}=2g_1\left(\partial^0A^{ij}-\partial^i\partial^j A^{0}\right)\equiv E^{ij}\ ,
\label{electricfielddef}\end{equation}
which indeed coincides with the tensor electric field \eqref{electricfielddefPretko}. Hence, finally, the answer to the question is the following: the electric tensor field $E^{ij}(x)$ \eqref{electricfielddefPretko} introduced in \cite{Pretko:2016lgv,Pretko:2016kxt,Pretko:2017xar,Du:2021pbc} is defined as the canonical momentum $\Pi_{ij}$ \eqref{Pi^ij} of the fracton action $S_{fract}$ \eqref{Sfract}, evaluated on the EoM \eqref{eom00F}.
In addition to the properties listed in Table \ref{table1}, which hold in general, the particular solution \eqref{A0} implies also
	\begin{empheq}{align}
	&F^{i00}=F^{0i0}=F^{00i}=0\label{F00i=0}\\
	&F^{ij0}=-2F^{0ij}=-2F^{i0j}\label{Fij0=-2F0ij}\ ,
	\end{empheq}
which hold for fractons only. As anticipated, because of \eqref{F00i=0}, for the fracton theory $g_2=0$ we have an additional Hamiltonian constraint, besides \eqref{Pi00=0}
\be
(\Pi^{i0}|_{g_2=0})|_\eqref{A0}=-g_1F^{i00}=0\ ,
\label{Pi00=0v2}\ee
which corresponds to the fact that, like in LG, also for fractons the degrees of freedom concern only the spatial components $A_{ij}(x)$.
Moreover, again in surprising analogy with Maxwell theory where the electric field and the field strength are related by $E^i(x)=-F^{0i}(x)$, we have that
\be
E^{ij}=-g_1 F^{ij0}=2g_1 F^{0ij}=2g_1 F^{i0j}\ .
\label{Eij}\ee
Taking \eqref{eom2} at $\alpha=0$ and $\beta=i$ (or, equivalently, \eqref{eomtot0i} at $g_2=0$) we have 
\begin{equation}\label{eom0i}
\partial_\mu F^{0i\mu} = \partial_jF^{0ij} = -\frac{1}{2}\partial_jF^{ij0} = 0\ ,
\ee
which, using \eqref{Eij}, writes
\be
\partial_jE^{ij}=0\ ,
\label{gauss-v}\ee
which is the vacuum Gauss law for the electric  tensor field \eqref{electricfielddef}. It is the tensorial extension of
\begin{equation}
		\vec\nabla\cdot\vec E=0\ .
		\end{equation}
Eq. \eqref{gauss-v} trivially implies
	\begin{equation}\label{gauss}
	\partial_i\partial_jE^{ij}=0\ ,
	\end{equation}
 which, together with \eqref{gauss-v}, is crucial for the property of limited mobility characterizing the fracton quasiparticles \cite{Nandkishore:2018sel,Pretko:2020cko,Pretko:2016lgv,Pretko:2016kxt}. As we shall show in a moment, while the Gauss-like equation \eqref{gauss-v} holds only for the fracton action \eqref{SfractF}, an equation formally identical to the limited mobility equation \eqref{gauss} holds for the LG action $S_{LG}$ \eqref{SLGF}, too.  
In fact, taking the divergence $\partial_i$ of the whole EoM \eqref{eomtot0i} and using \eqref{eomtot00} we have, at $g_1=0$, $i.e.$ for LG only,
\be
\partial_i\partial^iF_\lambda^{\ \lambda0} + 2 \partial_i\partial_jF^{ij0}=
-\partial_i\partial_j\Pi^{ij}|_{g_1=0}=0\ ,
\label{Piij=0LG}\ee
where we used the cyclicity property in Table \ref{table1} of the tensor $F_{\mu\nu\rho}(x)$, which in particular implies
\be
\partial_i\partial_jF^{0ij}=-\frac{1}{2}\partial_i\partial_jF^{ij0}\ .
\label{}\ee
The equation \eqref{Piij=0LG} is formally identical to its fracton counterpart \eqref{gauss}, and its possible consequences on the limited mobility of the gravitational waves are worth to investigate and to interpret.
Finally, taking \eqref{eom2} at $\alpha=i$ and $\beta=j$ (or, equivalently, \eqref{eomtotij} at $g_2=0$), we have
\begin{equation}\label{eomfractij}
\partial_\mu F^{ij\mu}
=\partial_0F^{ij0}+\partial_kF^{ijk}
=-\frac{1}{g_1}\partial_0E^{ij}+\partial_kF^{ijk}=0\ ,
\ee
where we used the definition of the electric tensor field $E^{ij}(x)$ \eqref{Eij}.
The fracton EoM \eqref{eomfractij} suggests to define the magnetic tensor field, in analogy  with the ordinary  vector magnetic field \mbox{$B_i(x)=\epsilon_{ijk}\partial^jA^k(x)=\frac{1}{2}\epsilon_{ijk}F^{jk}(x)$}, as
\begin{equation}\label{Bij}
		B_{i}^{\ j}\equiv g\epsilon_{0ilk}\partial^lA^{jk} = \frac{g}{3}\epsilon_{0ikl}F^{jkl}\ ,
\end{equation}
where $g$ is a constant to be suitably tuned. Its inverse is
\begin{equation}\label{B=F}
F^{ijk}\equiv-\frac{1}{g}\left(\epsilon^{0ikl}B^{\ j}_{l}+\epsilon^{0jkl}B^{\ i}_{l}\right)\ .
\end{equation}
The EoM \eqref{eomfractij} then can be written
\be
-\frac{1}{g_1}\partial_0E^{ij}-\frac{1}{g}\left(\epsilon^{0ikl}\partial_kB^{\ j}_{l}+\epsilon^{0jkl}\partial_kB^{\ i}_{l}\right)=0\ ,
\label{ampere}\end{equation}
which turns out to be completely analogous to the electromagnetic Amp\`ere law of electromagnetism in vacuum
		\begin{equation}
		-\partial_t\vec E+\vec\nabla\times\vec B=0\ ,
		\end{equation}
    of which \eqref{ampere} is the tensorial extension. It coincides with Eq.(26) in \cite{Pretko:2016lgv}.
From the definition \eqref{Bij} we find that the magnetic tensor field is traceless
\be
	B^{\;p}_p=0\label{TrB=0}\ ,
\ee
	and satisfies
\be
	\partial^aB^{\;p}_a=0\ \label{DivB=0}\ ,
\ee
which  is analogous to the standard Maxwell equation		
\begin{equation}
		\vec\nabla\cdot\vec B=0\ ,
		\end{equation}
and coincides with Pretko's second equation (38) in \cite{Pretko:2016lgv}. As in standard electromagnetism, the equation \eqref{DivB=0} is a geometric property, consequence of the definition of the magnetic tensor field $B_i^{\ j}(x)$ \eqref{Bij}.\\

Let us now study which information comes from the ``Bianchi'' identity in Table \ref{table1}
\begin{itemize}
\item $\alpha=0,\beta=j$
\begin{equation}\label{divB=0}
\epsilon_{0iab}\partial^{i}F^{jab}=\frac{3}{g}\partial^iB_{i}^{\;j}= 0\ ,
\end{equation}
we therefore recover the tensor magnetic Gauss law \eqref{DivB=0}.
\item $\alpha=l,\beta=i$
\begin{equation}\label{faraday}
		\begin{split}
		0&=\epsilon_{l\mu\nu\rho}\partial^{\mu}F^{i\nu\rho}\\
		&=\epsilon_{l\mu j0}\partial^\mu F^{ij0}+\epsilon_{l\mu jk}\partial^\mu F^{ijk}+\epsilon_{l\mu 0j}\partial^\mu F^{i0j}\\
		&=\frac{3}{2}\epsilon_{lmj0}\partial^m F^{ij0}+\epsilon_{l0 jk}\partial^0 F^{ijk}\\
		&=\frac{3}{2g_1}\epsilon_{0lmj}\partial^mE^{ij}+\frac{3}{g}\partial_0B^{\;i}_{l}\ ,
		\end{split}
	\end{equation}
where we used \eqref{F00i=0}, \eqref{Fij0=-2F0ij}, \eqref{Eij} and the definition \eqref{Bij}.
Eq.\eqref{faraday} is new, and it is the tensorial extension of the Faraday equation of electromagnetism: 
\begin{equation}
		\vec\nabla\times\vec E+\partial_t\vec B=0\ .
\end{equation}
It coincides with Pretko's Eq.(36) in \cite{Pretko:2016lgv}. 
\item
$\alpha=\beta=0$ and $\alpha=i,\ \beta=0$ 
are trivial identities.
\end{itemize}

Summarizing, from the EoM \eqref{eom2} and the ``Bianchi'' identity in Table \ref{table1} we have the following strong analogy with classical electromagnetism :
	\begin{empheq}{align}
	\mbox{\bf Maxwell} \qquad& \mbox{\bf Fractons} \nonumber \\
	\vec\nabla\cdot\vec E=0\qquad&\partial_jE^{ij}=0\label{max1}\\
	\vec\nabla\cdot\vec B=0\qquad&\partial^aB^{\;p}_a=0\label{max2}\\
	\vec\nabla\times\vec E-\partial_t\vec B=0\qquad&\epsilon_{0lmj}\partial^mE^{ij}-\frac{2g_1}{g}\partial^0B^{\;i}_{l}=0\label{max3}\\
	\vec\nabla\times\vec B-\partial_t\vec E=0\qquad&-\partial_0E^{ij}-\frac{2g_1}{g}\left(\frac{\epsilon^{0ikl}\partial_kB^{\ j}_{l}+\epsilon^{0jkl}\partial_kB^{\ i}_{l}}{2}\right)=0\ .\label{max4}
	\end{empheq}

Setting 
	\begin{equation}\label{galpha1}
	\frac{2g_1}{g}=-1
	\end{equation} 
    the last two equations  \eqref{max3} and \eqref{max4} are fully analogous to the corresponding ordinary Maxwell equations at the left hand side, and coincides with those introduced by Pretko in \cite{Pretko:2016lgv} from a completely different point of view, where actually it is written \eqref{gauss} rather than the more fundamental \eqref{gauss-v}.

\subsection{Fracton action in terms of electric and magnetic tensor fields}

We have seen that the fracton action \eqref{Sfract}, originally written in terms of the field $A_{\mu\nu}(x)$, can be written in terms of the tensor $F_{\mu\nu\rho}(x)$ as \eqref{SfractF}. This makes apparent the strong analogy with the classical electromagnetic Maxwell theory, of which the fracton theory appears to be the higher rank generalization. This analogy is even more spectacular when the four equations \eqref{max1}, \eqref{max2}, \eqref{max3} and \eqref{max4} governing the theory are considered, which can be written in terms of the two electric and magnetic tensor fields $E^{ij}(x)$ \eqref{electricfielddef} and $B_i^{\ j}(x)$ \eqref{Bij}.
As in Maxwell theory, two equations, namely \eqref{max1} and \eqref{max4}, are the EoM of the action \eqref{SfractF}, while the other two, \eqref{max2} and \eqref{max3}, are consequences of the ``Bianchi'' identity written in Table \ref{table1} for the tensor $F_{\mu\nu\rho}(x)$, hence have a geometrical nature. The analogy with electromagnetism can be pushed further by noting that the fracton action \eqref{SfractF} can be written in terms of the electric and magnetic tensor fields as follows:
\begin{equation}\label{actionE,B}
		\begin{split}
		S_{fract}&
		= {\frac{g_1}{6}}\int d^4x\,F_{\mu\nu\rho}F^{\mu\nu\rho}\\
		&= {\frac{g_1}{6}}\int d^4x\left(F^{0ij}F_{0ij}+F^{ij0}F_{ij0}+F^{i0j}F_{i0j}+F^{ijk}F_{ijk}\right)\\
		&= {\frac{g_1}{6}}\int d^4x\left(\frac{1}{4}F^{ij0}F_{ij0}+F^{ij0}F_{ij0}+\frac{1}{4}F^{ij0}F_{ij0}+F^{ijk}F_{ijk}\right)\\
		&= {\frac{g_1}{6}}\int d^4x\left(\frac{3}{2}F^{ij0}F_{ij0}+F^{ijk}F_{ijk}\right)\\
		&= {\frac{g_1}{6}}\int d^4x\left[-\frac{3}{2} {\frac{1}{g_1^2}}E^{ij}E_{ij}- {\frac{2}{g^2}\epsilon^{0kmn}B^{\;l}_{n}\epsilon_{0kab}B_{\;c}^{b}\left(\delta^a_m\delta^c_l+\delta^a_l\delta^c_m\right)}\right]\\
		&=\int d^4x\left(- {\frac{1}{4g_1}}E^{ij}E_{ij}+ {\frac{g_1}{g^2}}B_{i}^{\;j}B^{i}_{\;j}\right)\\
		&=\frac{1}{2}\int d^4x\left(- {\frac{1}{2g_1}}E^{ij}E_{ij}+ {\frac{2g_1}{g^2}}B^{\;j}_{i}B_{\;j}^{i}\right)\ ,
		\end{split}
	\end{equation}
where, besides the definitions of the electric and magnetic tensor fields \eqref{electricfielddef} and \eqref{Bij}, we used the properties of $F_{\mu\nu\rho}(x)$ \eqref{F00i=0} and \eqref{Fij0=-2F0ij}, and the tracelessness of the tensor magnetic field \eqref{TrB=0}. The result \eqref{actionE,B} closely reminds the electromagnetic action, whose Lagrangian is proportional to $E^2-B^2$, provided that
\begin{equation}\label{galpha}
	g^2=4g_1^2\ ,
	\end{equation}
which is compatible with the previously found constraint \eqref{galpha1}, that we will assume from now on.

\subsection{Stress-energy tensor and conservation laws}

The stress-energy tensor for the fracton action $S_{fract}$ \eqref{SfractF} is
	\begin{equation}\label{Tmunu}
		\begin{split}
		T_{\alpha\beta}&=\left.-\frac{2}{\sqrt{-g}}\frac{\delta S_{fract}}{\delta g^{\alpha\beta}}
		\right|_{g^{\alpha\beta}=\eta^{\alpha\beta}}\\
		&=\left.-\frac{ {g_1}}{3\sqrt{-g}}\frac{\delta }{\delta g^{\alpha\beta}}\int d^4x\sqrt{-g}g^{\mu\lambda}g^{\nu\gamma}g^{\rho\sigma}F_{\lambda\gamma\sigma}F_{\mu\nu\rho}
		\right|_{g^{\alpha\beta}=\eta^{\alpha\beta}}\\
		&= {\frac{g_1}{6}}\eta_{\alpha\beta}F^2-\frac{g_1}{3}\eta_{\alpha\gamma}\eta_{\beta\lambda}\left(2F^{\lambda\nu\rho}F_{\ \,\nu\rho}^\gamma+F^{\mu\nu\lambda}F_{\mu\nu}^{\ \ \gamma}\right)\ .
		\end{split}
	\end{equation}
Notice that taking the trace of $T_{\mu\nu}$, we have, in $d$-spacetime dimensions
\begin{equation}
	T=\eta^{\alpha\beta}T_{\alpha\beta}=\left.g_1\frac{(d-6)}{6}F^2\right|_{d=4}=-\frac{g_1}{3}F^2\ ,
	\label{traceTmunu}\end{equation}
which does not vanish in $d=4$, differently from what happens in Maxwell theory. The tracelessness of the stress-energy tensor would be recovered in $d=6$, which, as already remarked in the Introduction, seem to be the most natural, although unphysical, spacetime dimensions for fractons. The non-vanishing of the trace of the fracton stress-energy tensor is the sign that the theory, already at classical level, is not scale invariant. This suggests the existence of an energy scale. Now, since tracelessness is eventually related to the masslessness of the photon, the fact that the trace \eqref{traceTmunu} does not vanish might suggest the existence of a mass (as the typical energy scale) for the fractons, which can be introduced in a similar way as in LG \cite{Blasi:2017pkk,Blasi:2015lrg}.
The components of the stress-energy tensor are physically interpretable as follows~:
	\begin{itemize}
	\item $\alpha=\beta=0$ gives the energy density $T_{00}=u$  
		\begin{equation}\label{T00}
			\begin{split}
T_{00}=u
&=
- \frac{g_1}{6}F^2+\frac{g_1}{3}\left(2F^{0\mu\nu}F_{0\mu\nu}+F^{\mu\nu0}F_{\mu\nu0}\right)
\\
&=- 
\frac{g_1}{6}F^2+\frac{g_1}{2}F^{ij0}F_{ij0}
\\
&=
\frac{1}{4g_1}\left( E^{ij}E_{ij}- B^{\;j}_iB_{\;j}^i\right)+\frac{g_1}{2}\left(- {\frac{1}{g_1}}\right)\left( {\frac{1}{g_1}}\right)E^{ij}E_{ij}
\\
&=
-\frac{1}{4g_1}\left(E^{ij}E_{ij}+B^{\;j}_iB_{\;j}^i\right)\ ,
\end{split}
\end{equation}
where \eqref{F00i=0}, \eqref{Fij0=-2F0ij}, \eqref{actionE,B}, \eqref{Eij} and \eqref{galpha} have been used. Again, this expression is formally identical to the corresponding electromagnetic result $u\propto E^2+B^2$.
From the positivity constraint of the energy density $u$ it  must be
		\begin{equation}
		g_1<0\ ,
		\end{equation}
and, from now on, we choose
\be
g_1=-1\ ;
\label{g1=-1}\ee
\item $\alpha=0,\ \beta=i$ gives the Poynting vector $T_{0i}=S_i$ 	
\begin{equation}
\begin{split}
T_{0i}=S_i=
& 
\frac{1}{3}\eta_{i\lambda}\left(2F^{\lambda\nu\rho}F_{0\nu\rho}+F^{\mu\nu\lambda}F_{\mu\nu0}\right)
\\
=& 
\frac{1}{3}\eta_{ij}\left(2F^{jkl}F_{0kl}+F^{klj}F_{kl0}\right)
\\
=&
\frac{1}{6}\eta_{ij}\left[-\left(\epsilon^{0jlp}B_p^{\;k}+\epsilon^{0klp}B_p^{\;j}\right)E_{kl}+\left(\epsilon^{0kjp}B_p^{\;l}+\epsilon^{0ljp}B_p^{\;k}\right)E_{kl}\right]
\\
=&
\frac{1}{6}\eta_{ij}E_{kl}\left(-2\epsilon^{0jlp}B_p^{\;k}-\cancel{\epsilon^{0klp}B_p^{\;j}}+\epsilon^{0kjp}B_p^{\;l}\right)
\\
=&
-\frac{1}{2}\eta_{ij}E_{kl}\epsilon^{0jlp}B_p^{\;k}
\\
=&\frac{1}{2}\epsilon_{0ilp}E^{kl}B^p_{\;k}\ ,
\end{split}
\end{equation}
which, as in Maxwell electromagnetism, is the vector product of the electric and magnetic tensor fields $\vec S\propto\vec E\times\vec B$\ ;
\item $\alpha=i,\ \beta=j$ gives the stress tensor $T_{ij}=\sigma_{ij}$ 
\begin{equation}
\begin{split}
T_{ij}&=
-\frac{1}{6}\eta_{ij}F^2+\frac{1}{3}\eta_{jk}\left(2F^{k\mu\nu}F_{i\mu\nu}+F^{\mu\nu k}F_{\mu\nu i}\right)
\\
&=
-\frac{1}{6}\eta_{ij}F^2+\frac{1}{3}\eta_{jk}\left(3F^{ka0}F_{ia0}+2F^{kab}F_{iab}+F^{abk}F_{abi}\right)
\\
&=
-\frac{1}{6}\eta_{ij}F^2-\eta_{jk}E^{ka}E_{ia}+\frac{1}{3}\eta_{jk}\left(2F^{kab}F_{iab}+F^{abk}F_{abi}\right)
\\
&=
-\frac{1}{6}\eta_{ij}F^2-\eta_{jk}E^{ka}E_{ia}+\frac{1}{6}\eta_{jk}\left(2\delta^k_i B_a^{\;b}B^a_{\;b}-2B_i^{\;a}B^k_{\;a}+4B_a^{\;k}B^a_{\;i}-\epsilon^{0akp}\epsilon_{0biq}B_p^{\;b}B^q_{\;a}\right)
\\	
&=
-\frac{1}{6}\eta_{ij}F^2-\eta_{jk}E^{ka}E_{ia}+\frac{1}{2}\eta_{jk}\left(\delta^k_i B_a^{\;b}B^a_{\;b}-B_i^{\;a}B^k_{\;a}+B_a^{\;k}B^a_{\;i}\right)
\\	
&=
\eta_{ij}T_{00}-\eta_{jk}\eta_{il}E^{ka}E^l_{\;a}-\frac{1}{2}\eta_{jk}\eta_{il}\left(B^{la}B^k_{\;a}-B_a^{\;k}B^{al}\right)\ ,
\end{split}
\end{equation}
where we used
\begin{empheq}{align}
2F^{kab}F_{iab}
&=
-\frac{1}{2}\left(\epsilon^{0kbp}B_p^{\;a}+\epsilon^{0abp}B_p^{\;k}\right)\left(\epsilon_{0ibq}B^q_{\;a}+\epsilon_{0abq}B^q_{\;i}\right)
\\
&=
\frac{1}{2}\left(\delta^k_i B_a^{\;b}B^a_{\;b}-B_i^{\;a}B^k_{\;a}+4B_a^{\;k}B^a_{\;i}\right)\nonumber\\
F^{abk}F_{abi}
&=
-\frac{1}{4}\left(\epsilon^{0akp}B_p^{\;b}+\epsilon^{0bkp}B_p^{\;a}\right)\left(\epsilon_{0aiq}B^q_{\;b}+\epsilon_{0biq}B^q_{\;a}\right)
\\
&=
\frac{1}{2}\left(\delta^k_i B_a^{\;b}B^a_{\;b}-B_i^{\;a}B^k_{\;a}-\epsilon^{0akp}\epsilon_{0biq}B_p^{\;b}B^q_{\;a}\right)\ ,
\nonumber
\end{empheq}
and, from \cite{landau}, 
		\begin{equation}\label{epsxeps}
		\epsilon^{0akp}\epsilon_{0biq}=-\delta^a_b(\delta^k_i\delta^p_q-\delta^p_i\delta^k_q)+\delta^a_i(\delta^k_b\delta^p_q-\delta^p_b\delta^k_q)-\delta^a_q(\delta^k_b\delta^p_i-\delta^p_b\delta^k_i)\ ,	
		\end{equation}
so that
	\begin{equation}
	\epsilon^{0akp}\epsilon_{0biq}B_p^{\;b}B^q_{\;a}=-\delta^k_iB_a^{\;b}B^a_{\;b}+B_i^{\;a}B_{\;a}^k+B_a^{\;k}B^a_{\;i}\ ,
	\end{equation}
because of the tracelessness of the magnetic tensor \eqref{TrB=0}. 
Finally, we used also the fact that, because of 
\eqref{actionE,B} and \eqref{T00} we have  \mbox{$T_{00}=-\frac{1}{6}F^2+\frac{1}{2} B_a^{\;b}B^a_{\;b}$}.
As expected, the stress tensor is symmetric $i\leftrightarrow j$~:
\begin{equation}
\begin{split}
T_{ij}=
\frac{1}{4}\eta_{ij}
\left(
E^{ab}E_{ab}+B^{\;b}_aB_{\;b}^a
\right)
-\eta^{ab}E_{ia}E_{jb}-\frac{1}{2}\eta^{ab}\left(B_{ia}B_{jb}-B_{aj}B_{bi}\right)\ .
\end{split}
\end{equation}
Once again, the analogy with Maxwell theory, for which the stress tensor is
	\begin{equation}
	\sigma_{ij}=\frac{1}{2}\eta_{ij}\left(E^2+B^2\right)-E_iE_j-B_iB_j\ ,
	\end{equation}
is impressive.
\end{itemize}
Let us now discuss the (on-shell) conservation of the stress-energy tensor 
\be
\partial^\nu T_{\mu\nu} =0\ ,
\label{partialTmunu=0}\ee
whose components are
\begin{itemize}
\item $\mu=0$:
\begin{equation}\label{F0}
\begin{split}
\partial^\nu T_{\nu0}
&=
\partial^0T_{00}+\partial^iT_{i0}
\\
&=
\partial^0u+\partial^iS_i
\\
&=
-\frac{1}{4}\partial_0\left(E_{ab}E^{ab}+B_a^{\;b}B^a_{\;b}\right)
+\frac{1}{2}\epsilon_{0ilp}\partial^i\left(E^{kl}B_{\;k}^p\right)
\\
&=
-\frac{1}{2}\left[ E_{ab}\partial_0E^{ab}+B^a_{\;b}\partial_0B_a^{\;b}
-\epsilon_{0ilp}\partial^i\left(E^{kl}B_{\;k}^p\right)\right]
\\ 
&=-\frac{1}{2}\left[
\epsilon^{0akl}E_{ab}\partial_kB_l^{\;b}+\epsilon_{0amn}B^a_{\;b}\partial^mE^{bn}
-\epsilon_{0ilp}\partial^i\left(E^{kl}B_{\;k}^p\right)\right]
\\
&=
\frac{1}{2}\epsilon_{0amn}\left[
E^{ab}\partial^mB^n_{\;b}+B^n_{\;b}\partial^mE^{ab}
-\partial^m\left(E^{ab}B^n_{\;b}\right)\right]
\\
&=0\ ,
\end{split}
\end{equation}
where we used the EoM \eqref{ampere} and \eqref{faraday}. The continuity equation is therefore verified on-shell
\begin{equation}
	\partial^iS_i+\partial^0u=0\ .
\label{continuity}\end{equation}

\item $\mu=i$:
\begin{equation}\label{Fi}
\begin{split}	
\partial^\nu T_{\nu i}&=\partial^0T_{0i}+\partial^jT_{ji}\\
&=
\partial^0S_i+\partial^j\sigma_{ji}
\\
&=
-\frac{1}{2}\epsilon_{0imn}\partial_0\left(E^{am}B^n_{\;a}\right)+\partial_iu-\eta_{il}\partial_k\left(E^{ka}E^l_{\;a}\right)-\eta_{il}\partial_k\left(B^{la}B^k_{\;a}-B_a^{\;k}B^{al}\right)
\\
&=
-\frac{1}{2}\epsilon_{0imn}\left\{\left[\frac{1}{2}\left(\epsilon^{0akl}\partial_kB_l^{\;m}+\epsilon^{0mkl}\partial_kB_l^{\;a}\right)\right]B^n_{\;a}-\epsilon^{0nbc}\partial_bE_{ac}E^{am}\right\}+
\\
&\quad
+\partial_iu-\eta_{il}\partial_k\left(E^{ka}E^l_{\;a}\right)-\eta_{il}\partial_k\left(B^{la}B^k_{\;a}-B_a^{\;k}B^{al}\right)
\\
&=
-\frac{1}{4}\left[\epsilon_{0imn}\epsilon^{0akl}\partial_kB_l^{\;m}+\left(\delta^k_i\delta^l_n-\delta^k_n\delta^l_i\right)\partial_kB_l^{\;a}\right]B^n_{\;a}+
\\
&\quad
-\frac{1}{2}\left(\delta^b_i\delta^c_m-\delta^b_m\delta^c_i\right)E^{am}\partial_bE_{ac}+\partial_iu-\eta_{il}\partial_k\left(E^{ka}E^l_{\;a}\right)-\eta_{il}\partial_k\left(B^{la}B^k_{\;a}-B_a^{\;k}B^{al}\right)
\\
&=
-\frac{1}{4}\left(-B^n_{\;i}\partial_mB_n^{\;m}-2B^n_{\;m}\partial_nB_i^{\;m}+\cancel{2B^n_{\;m}\partial_iB_n^{\;m}}\right)+
\\
&\quad
-\frac{1}{2}\left(\cancel{E^{ac}\partial_iE_{ac}}-E^{ab}\partial_bE_{ai}\right)+\cancel{\partial_iu}-E_{ia}\partial_kE^{ka}-E^{ka}\partial_kE_{ia}+
\\
&\quad
-\frac{1}{2}\left(B^n_{\;m}\partial_nB_i^{\;m}-B^n_{\;i}\partial_mB_n^{\;m}-B_n^{\;m}\partial_mB^n_{\;i}\right)
\\
&=
\frac{1}{4}\left[
\left(3B^n_{\;i}\partial_mB_n^{\;m}+2B_n^{\;m}\partial_mB^n_{\;i}\right)-2E^{ab}\partial_bE_{ai}\right]\\
&\neq 0\ ,
\end{split}
\end{equation}	
where \eqref{ampere}, \eqref{faraday}, \eqref{epsxeps} and \eqref{gauss-v} have been used. Differently from the continuity equation \eqref{continuity}, the spatial components of the divergence of the stress-energy tensor do not vanish. Now, what should we expect actually ? If we think of the stress-energy tensor as the conserved current associated to the diffeomorphism symmetry, as the definition \eqref{Tmunu} suggests, it should not be conserved in a theory like the fracton one, which is not diffeomorphism invariant \eqref{notdiff}. On the other hand, we are facing here with a partial conservation of the stress-energy tensor, because its time component is indeed conserved, yielding the continuity equation \eqref{continuity}, which relates the flux of the energy density to the divergence of the momentum density.  The partial conservation of the stress-energy tensor might be explained by observing that the fracton symmetry \eqref{dA} is indeed a diff transformation \eqref{diff} with a particular choice of the vector diff parameter.
\end{itemize}

\section{Fracton Lorentz force}

It is interesting to study how the physics is modified if matter is introduced by means of a symmetric rank-2 tensor $J^{\mu\nu}(x)=J^{\nu\mu}(x)$ coupled to the fracton field $A_{\mu\nu}(x)$
\be
S_{fract}\rightarrow S_{tot}=S_{fract}+S_J\ ,
\label{Stot}\ee
where $S_{fract}$ is the pure fractonic action \eqref{Sfract} (or \eqref{SfractF}), and $S_J$ is the matter action
\begin{equation}
	S_J\equiv-\int d^4x\,J^{\mu\nu}A_{\mu\nu}\ .
\label{Smatter}\end{equation}
The EoM \eqref{eom2} modifies as
\begin{equation}\label{eomJ}
	\partial_\mu F^{\alpha\beta\mu}=-J^{\alpha\beta}\ .
\end{equation}
We observe that, due to the cyclicity identity in Table \ref{table1}, $J^{\alpha\beta}$ is conserved in the following sense
\begin{equation}\label{divJ=0}
	\partial_\alpha\partial_\beta J^{\alpha\beta}=0\ .
	\end{equation}
The components of the EoM \eqref{eomJ} are
	\begin{itemize}
	\item $\alpha=\beta=0$ :
	\begin{equation}\label{eomJ00}
	\partial_i F^{00i}=-J^{00}=0\ ,
	\end{equation}
which vanishes because of \eqref{F00i=0}, consequence of \eqref{A0}. Hence, there is no coupling with $A_{00}(x)$, as expected,  since it is not a dynamical degree of freedom of the theory, due to \eqref{Pi00=0}.

\item $\alpha=0,\ \beta=i$ :
	\begin{equation}\label{eomJ0i}
	\partial_j F^{0ij}=-J^{0i}\ ,
	\end{equation}
which, using \eqref{Eij}, becomes
	\begin{equation}\label{gauss-vett}
	\partial_jE^{ij}=2J^{i0}\ .
	\end{equation}
Taking the divergence of \eqref{gauss-vett} we find the analogous of the Gauss law 	\begin{equation}\label{gauss+mat}
	\partial_i\partial_jE^{ij}=\rho\ ,
	\end{equation}
where we defined the charge density
	\begin{equation}\label{rho}
	\rho\equiv2\partial_iJ^{i0} \ .
	\end{equation}
This equation plays a central role in \cite{Nandkishore:2018sel,Pretko:2020cko,Pretko:2016lgv,Pretko:2016kxt,Pretko:2017xar}, since it yields not only the charge neutrality condition, but also the vanishing of the total dipole moment. In fact, integrating \eqref{gauss+mat}, we get
	\begin{equation}
	\int dV\partial_i\partial_jE^{ij}=\int dV\,\rho=0\ ,
	\end{equation}
which states that the total charge inside an infinite volume is zero. Moreover, from \eqref{gauss+mat} we also have
	\begin{equation}\label{dipole cons}
	\int dVx^k\partial_i\partial_jE^{ij}=
	\int dV\,x^k\rho=\int dV\,p^k=0\ ,
	\end{equation}
according to which the dipole moment density, defined as 
\be
p^k=x^k\rho\ ,
\label{dipoledensity}\ee 
of an infinite volume vanishes.

\item $\alpha=i,\ \beta=j$ :
	\begin{equation}\label{eomJij}
	\partial_\mu F^{ij\mu}=-J^{ij}\ ,
	\end{equation}
which, using \eqref{ampere}, becomes
	\begin{equation}\label{eomJij2}
	-\partial_0E^{ij}+\frac{1}{2}\left(\epsilon^{0ikl}\partial_kB^{\ j}_{l}+\epsilon^{0jkl}\partial_kB^{\ i}_{l}\right)=J^{ij}\ .
	\end{equation}
Differentiating 
 with $\partial_i\partial_j$, we have
	\begin{equation}\label{}
		\begin{split}
		0&=\partial_0\partial_i\partial_jE^{ij}+\partial_i\partial_jJ^{ij}\\
		&=\partial_0\rho+\partial_i\partial_jJ^{ij}\ ,
		\end{split}
	\end{equation}
    where \eqref{gauss+mat} has been used. It is a kind of continuity equation \cite{Pretko:2016lgv,Du:2021pbc,Jain:2021ibh}, which can also be obtained from the conservation equation \eqref{divJ=0}
	\begin{equation}
	0=\partial_\alpha\partial_\beta J^{\alpha\beta}=2\partial_0\partial_iJ^{0i}+\partial_i\partial_jJ^{ij}=\partial_0\rho+\partial_i\partial_jJ^{ij}\ ,
	\end{equation}
	where we have used the definition of the density charge $\rho(x)$ \eqref{rho} and the fact that $J^{00}(x)=0$ \eqref{eomJ00}.
	\end{itemize}
It is also interesting to see how the (partial) conservation of the stress-energy tensor is modified by the presence of matter. 
The continuity equation \eqref{continuity} is modified as
\be
\partial^\nu T_{\nu0}
=
\partial^0T_{00}+\partial^iT_{i0}=
\partial^0u+\partial^iS_i= E_{ab}J^{ab}\ ,
\label{continuitymatter}\ee
while the spatial components of \eqref{Fi} acquire the term in the last row
\bea
\partial^\nu T_{\nu i}=\partial^0T_{0i}+\partial^jT_{ji}
=
\partial^0S_i+\partial^j\sigma_{ji} &=&
\frac{1}{4}\left[
\left(3B^n_{\;i}\partial_mB_n^{\;m}+2B_n^{\;m}\partial_mB^n_{\;i}\right)-2E^{ab}\partial_bE_{ai}\right]\nonumber\\
&&+
\frac{1}{2}\epsilon_{0imn}J^{am}B^n_{\ a}-2J^{a0}E_{ia}\ ,
\label{Fimatter}\eea
The additional terms appearing in \eqref{continuitymatter} and \eqref{Fimatter} can be easily interpreted if, again, we think to the standard Maxwell theory of electromagnetism, where the divergence of the stress-energy tensor in presence of matter involves the 4D ``Lorentz force'' per unit volume on matter $f^\mu$: 
\be
\partial_\nu T^{\mu\nu}+f^\mu=0\ .
\label{divTmatter}\ee
At the right hand side of \eqref{continuitymatter} appears
\be
f^0=E_{ab}J^{ab}\ ,
\label{power}\ee
which is the analogous of the electromagnetic power $\vec{E}\cdot\vec{J}$. The last term at the right hand side of \eqref{Fimatter}
\be
f^i=  2J_{a0}E^{ia}-\frac{1}{2}\epsilon^{0imn}J_{am}B_n^{\ a}
\label{lorentzforce}\ee
can be traced back to the generalized Lorentz force on a dipole $p^i(x)$ moving with velocity $v^i$ proposed in \cite{Pretko:2016lgv}
\be
F^i=-p_jE^{ij}-\epsilon^{ilk}p_jv_lB_k^{\ j}\ ,
\label{pretkoloretnzforce}\ee
once we take
\be
J^{0i}\sim p^i\ ,
\label{J0ipi}\ee
and
\be
J^{ij}\sim p^iv^j+p^jv^i\ .
\label{Jijpivj}\ee
The first identification \eqref{J0ipi} is compatible with \eqref{dipole cons}, when \eqref{gauss-vett} is taken into account, and the second relation \eqref{Jijpivj} agrees with the microscopic lattice definition of the current of a dipole made in \cite{Pretko:2016lgv,Xu:2006}. What is remarkable is that we recover here as part of the conservation law of the stress-energy tensor the picture conjectured in \cite{Pretko:2016lgv}: the isolated electric monopoles of the theory described by the action \eqref{SfractF} are fractons, which do not respond to the electromagnetic fields and, hence, do not move, due to the dipole conservation constraint \eqref{gauss+mat}. What we find here is that, instead, dipole motion, which preserves the global dipole moment, does respond to the electromagnetic field tensors $E^{ij}(x)$ and $B_i^{\ j}(x)$ according to \eqref{lorentzforce}, like a conventional charge particle responds to an ordinary electromagnetic field. Hence, from \eqref{lorentzforce}, we confirm the ``intuition'' proposed in \cite{Pretko:2016lgv} concerning the Lorentz force on a fracton dipole \eqref{pretkoloretnzforce}.
\normalcolor

\section{$\theta$-term}

In ordinary vector gauge field theory, it is known that a term can be added to the Maxwell action (or to its non-abelian extension, namely the Yang-Mills theory): the so called $\theta$-term, which has the form
\be
S_\theta\sim\theta\int d^4x\;\epsilon^{\mu\nu\rho\sigma}F_{\mu\nu}F_{\rho\sigma}\sim
\theta\int d^4x\;\vec{E}\cdot\vec{B}\ ,
\label{Stheta}\ee
where $\theta$ is a constant parameter. The $\theta$-term represented by \eqref{Stheta} is topological, since it does not depend on the spacetime metric, and it is a total derivative, hence it does not contribute to the EoM. Nonetheless, the $\theta$-term is relevant in several contexts, like axion electrodynamics, the Witten effect and the strong CP problem (see for instance \cite{tong,Peccei:1977hh,Peccei:1977ur,Sikivie:1983ip,Wilczek:1987mv}). 
For what concerns the fracton theory,  in \cite{Pretko:2017xar} the $\theta$-term has been generalized as $E_{ij}B^{ij}$, in analogy with the fractonic Hamiltonian density, assumed to be proportional to $(E_{ij}E^{ij}+B_{ij}B^{ij})$.
Considering a compact tensor gauge field, which implies a ``magnetic'' monopole $(\partial_iB^{ij}=g^i\neq0)$, the introduction of the $\theta$-term gives to the Gauss constraint an additional contribution related to the ``magnetic'' field. As for dyons in the Witten effect \cite{Witten:1979ey}, the ``electric'' charge gains an additional contribution related to the ``magnetic'' vector charge \cite{Nandkishore:2018sel,Pretko:2020cko,Pretko:2017xar}. On the other hand  the possibility of a non-constant $\theta$-term has not yet been investigated in the context of fractons. 
The motivation for such a generalization comes from the Topological Insulators, which are characterized by a step function $\theta$-term, which switches between $\theta=0$ outside the material and $\theta=\pi$ inside. Another example is given by
axion models \cite{Peccei:1977hh,Peccei:1977ur}, which 
describe a dynamical field coupled to photons via a local $\theta(x)$-term. This generates modified Maxwell equations \cite{Sikivie:1983ip,Wilczek:1987mv} as follows
	\begin{empheq}{align}
	\vec\nabla\cdot\vec E&=\rho-\vec\nabla\theta\cdot\vec B\label{axion1}\\
	\vec\nabla\times\vec B-\partial_t\vec E&=\vec J+\partial_t\theta\,\vec B+\vec \nabla\theta\times\vec E\ ,\label{axion2}
	\end{empheq}
where the additional terms contribute as an excess of charge ($\vec\nabla\theta\cdot\vec B$) and current ($\partial_t\theta\,\vec B+\vec \nabla\theta\times\vec E$) densities. Thus axion models are frequently used  in the context of condensed matter and Topological Insulators to mimic a non-constant $\theta$-term, like for instance in \cite{Rosenberg:2010ia}. As we are dealing with a rank-2 tensor theory, an interesting example is the case studied in \cite{Chatzistavrakidis:2020wum} in the context of LG, where modified gravitoelectromagnetic \cite{Mashhoon:2003ax} equations analogous to \eqref{axion1} and \eqref{axion2} are recovered.
According to the formalism presented in this article, based on the Maxwell-like construction of a consistent theory for fractons, the analogous of the $\theta$-term should be the following
\begin{equation}
		S_\theta=\frac{1}{9}\int d^4x\,\theta\epsilon_{\mu\nu\rho\sigma}F^{\lambda\mu\nu} F_\lambda^{\;\rho\sigma}\\
		=\int d^4x\,\theta\epsilon_{\mu\nu\rho\sigma}\partial^\mu A^{\lambda\nu}\partial^\rho A_{\lambda}^\sigma\ .
\label{Sthetafract}	\end{equation}
    We shall see that this is indeed the case by comparing the consequences of adding this term to the action $S_{fract}$ \eqref{Sfract} to the known results concerning the $\theta$-term in the theory of fractons \cite{Pretko:2017xar,Nandkishore:2018sel,Pretko:2020cko}
    and of LG \cite{Chatzistavrakidis:2020wum}. Notice that,  differently from the standard $\theta$-term introduced to solve the strong CP problem \cite{tong,Peccei:2006as}, $S_\theta$ is not topological, due to the contraction of the $\lambda$-indices in \eqref{Stheta}. Moreover, here $\theta(x)$ is not constant, like in \cite{Chatzistavrakidis:2020wum}. For instance, $\theta(x)$ might be the Heaviside step function, which  would correspond to introducing a boundary at $x=0$ \cite{Amoretti:2014iza,Bertolini:2020hgr,Bertolini:2021iku,Amoretti:2013xya,Bertolini:2022sao}. The contribution of $S_\theta$ to the EoM is
\begin{equation}\label{eomSt}
	\frac{\delta S_\theta}{\delta A^{\alpha\beta}}=-(\delta^\gamma_\alpha\delta^\sigma_\beta+\delta^\sigma_\alpha\delta^\gamma_\beta)\partial^\rho\theta\,\epsilon_{\mu\nu\rho\sigma}\eta_{\lambda\gamma}\partial^\mu A^{\lambda\nu}\ .
	\end{equation}
It is interesting to observe that this contribution is the same as the one that in \cite{Chatzistavrakidis:2020wum} gives the $\theta$-modified term of the gravitoelectromagnetic equations, where the electric and magnetic fields are vectors.
The EoM \eqref{eom2} acquire an additional term
\begin{equation}\label{eomS+St}
	\frac{\delta S_{fract}}{\delta A^{\alpha\beta}}+\frac{\delta S_\theta}{\delta A^{\alpha\beta}}=0\ ,
	\end{equation}
whose components are
\begin{itemize}
\item $\alpha=\beta=0$
		\begin{equation}
		\partial_iF^{00i}-2\partial^k\theta\,\epsilon_{0ijk}\partial^iA^{0j}=0\ ,
		\end{equation}
which is still solved by $A^{0\mu}=\partial^\mu A^0$ \eqref{A0}~;

\item $\alpha=0,\ \beta=i$ 
	\begin{equation}\label{eomt0i}
	\partial_jE^{ij}-\partial^j\theta\, B_j^{\ i}=0\ ,
	\end{equation}
which is the tensorial extension of \eqref{axion1}
	\begin{equation}
	\vec\nabla\cdot\vec E=-\vec\nabla\theta\cdot\vec B~;
	\end{equation}
	
\item $\alpha=i,\ \beta=j$ 
\begin{equation}
	\delta_{ab}^{ij}\left[\partial_0E^{ab}-\epsilon^{0akl}\partial_kB_l^{\;b}+\eta^{ac}\left(\epsilon^{0nm b}\partial_n\theta\, E_{mc}+\partial_0\theta\,B^b_{\;c}\right)\right]=0\ ,
	\end{equation}
where we defined the symmetrized delta 
\be
\delta^{ab}_{ij}\equiv\frac{1}{2}(\delta^a_i\delta^b_j+\delta^b_i\delta^a_j)\ ,
\label{}\ee
which agrees with \eqref{axion2} 	
\begin {equation}
	\vec\nabla\times\vec B-\partial_t\vec E=\partial_t\theta\,\vec B+\vec \nabla\theta\times\vec E\ .
	\end{equation}
\end{itemize}
Therefore Eqs. \eqref{eomS+St} are generalized tensorial $\theta$-modified Maxwell equations. In particular, as in the standard modified Maxwell equations \cite{Wilczek:1987mv}, we can interpret the $\theta$-dependent terms as an excess of charge and current densities: 
	\begin{equation*}
	\partial_{j}E^{ij}=\tilde\rho^i\quad;\quad	-\partial_0E^{ij}+\frac{1}{2}\left(\epsilon^{0ikl}\partial_kB^{\ j}_{l}+\epsilon^{0jkl}\partial_kB^{\ i}_{l}\right)=\tilde J^{ij}\ ,
	\end{equation*}
with
	\begin{equation*}
	\tilde\rho^i\equiv\partial^j\theta\, B_j^{\ i}\quad;\quad\tilde J^{ij}\equiv\delta_{ab}^{ij}\eta^{ac}\left(\epsilon^{0nm b}\partial_n\theta\, E_{mc}+\partial_0\theta\,B^b_{\;c}\right)\ .
	\end{equation*}
\normalcolor
We have a further confirmation that $S_\theta$ \eqref{Stheta} is indeed the correct $\theta$-term when we write it in terms of the electric and magnetic tensor fields \eqref{electricfielddef} and \eqref{Bij} 
\begin{equation}
S_\theta =
	-\frac{1}{3}\int d^4x\,\theta\eta_{lm}\epsilon_{0ijk}F^{il0}F^{mjk}
	=-\frac{1}{2}\int d^4x\,\theta E^{il}B_{il}\ ,
\label{}\end{equation}
which  is the tensorial extension of the standard $\theta$-term $S_\theta\sim\theta\int \vec{E}\cdot\vec{B}$ for  constant $\theta$ \cite{tong}.


\section{Final remarks}

In this paper we adopted a covariant approach to the theory of fractons. This is not only a matter of formalism, but, rather, it allows to better understand the nature itself of these quasiparticles. In the usual approach of the theory of fractons, space and time are treated separately, hence non covariantly. The standard way to proceed is to take the {\it spatial} Gauss contraint \eqref{dedeE=0}, written for a ``tensor electric field'' $E_{ij}(x)$,  as the tool to realize the defining property of fractons, $i.e.$ their limited mobility, by extending to fracton dipoles the usual conservation law holding for electric charges. This is usually achieved by introducing a ``tensor'', instead of a vector, potential $A_{ij}(x)$, obeying the generalized $spatial$ gauge transformation \eqref{deA=dedelambda}. As we explained, the tensor field $E_{ij}(x)$ was, somehow, defined as the $spatial$ canonical momentum, as in \eqref{electricfielddefPretko}. We say $somehow$ because in the definition of $E_{ij}(x)$ a scalar field $A_0(x)$ appears, as a multiplier introduced by hand in order to enforce the Gauss constraint.
Moreover, given the fracton limited mobility, it comes naturally to ask which is the generalization of the Lorentz force, and how the absence of motion could be compatible with the existence of an electromagnetic Lorentz force, and, above all, which is the elementary object on which such a force acts. Last but not least, in the Literature fractons are often seen in relation to gravity in a nontrivial way, starting from \cite{Pretko:2017fbf}. The main contribution of this paper is to show that, embedding the usual spatial and non covariant theory of fractons in a more general {\it covariant gauge field theory}, everything goes to the right place naturally, without introducing by hand any external ingredient. Our unique and starting point, as usual in field theory, is the covariant transformation \eqref{fractonsymintro}, from which the most general $covariant$ invariant action \eqref{Sinvg1g2} is derived. To cite a few new results following this approach, the relation with linearized gravity appears immediately, since the action \eqref{Sinvg1g2} consists of two terms, one of which, namely \eqref{SLG}, just describes linearized gravity. The theory of fractons as ``emergent electromagnetism'', as often has been called, is evident from the beginning as well, once we defined the rank-three ``electromagnetic'' tensor field \eqref{Fmunurho} by means of which the fracton Lagrangian writes as $F^2$, just like Maxwell theory (from which the title of this paper). 
According to our field-theoretic point of view, the Gauss constraint is not an external constraint anymore, but turns out to be one of the equations of motion, the others formally coinciding with the Maxwell equations, which is mostly interesting, in our opinion. Finally, studying the conservation of the stress-energy tensor, we recovered the Lorentz force \eqref{lorentzforce}, exactly as it was correctly guessed in \cite{Pretko:2016lgv}, from which we see that, as one might expect, the force acts on fracton dipoles, and not on isolated charges, thus preserving the absence of mobility for isolated fractons. 

\section*{Acknowledgments}

We thank Alberto Blasi  and Giandomenico Palumbo for enlightening discussions. This work has been partially supported by the INFN Scientific Initiative GSS: ``Gauge Theory, Strings and Supergravity''. E.B. is supported by MIUR grant ``Dipartimenti di Eccellenza'' (100020-2018-SD-DIP-ECC\_001). 

\newpage


\end{document}